\documentclass[usenatbib,useAMS]{mn2e}
\title[{\em Chandra}, {\em XMM-Newton} and {\em HST} observations of M31 transient T14]{{\em Chandra}, {\em XMM-Newton} and {\em HST} observations of a transient in M31 with a possible asymmetric, precessing disk}

\usepackage{graphicx}
\usepackage{float}
\author[Barnard et al.]{R. Barnard$^{1}$, M.~R. Garcia$^{1}$,  and S.~S. Murray$^{1,2}$\\
$^1$Harvard-Smithsonian Center for Astrophysics, Cambridge, MA 02138, USA\\
$^2$Johns Hopkins University, Baltimore, MD, USA}
\def\aap{A\&A}
\def\apj{ApJ}
\def\mnras{MNRAS}
\def\apjl{ApJL}
\def\pasj{PASJ}
\def\araa{ARA\&A}
\def\pasp{PASP}
\def\apjs{ApJS}
\def\aj{AJ}

\begin{document}

 \date{Accepted . Received ; in original form}

\pagerange{\pageref{firstpage}--\pageref{lastpage}} \pubyear{2015}
\maketitle
\label{firstpage}

\begin{abstract}

The transient X-ray source CXOM31 004205.77+411330.43  exhibited several outbursts during our long-term monitoring campaign of $\sim$monthly observations of the M31 center with {\em Chandra}. However, the decay profile appears to be  unlike those observed from Galactic transients.  We followed up the 2011 outburst with two $\sim$B band  {\em HST}/ACS observations, one in outburst and the other in quiescence, and used difference imaging to search for a counterpart; this would  be dominated by re-processed X-ray emission from the disk. We found a counterpart with B = 28.21$\pm$0.16. An  {\em XMM-Newton} observation from a previous outburst yielded a spectrum that {  is} well described by an absorbed power law with absorption equivalent to { $\sim$2.6$\times 10^{21}$ H atom cm$^{-2}$} and photon index $\sim$1.8; {   the highest quality {\em Chandra} spectrum yielded only $\sim$130 counts, and  best fits consistent with the {\em XMM-Newton} results}. We calculated an absolute V magnitude of { +1.9}  during outburst for a typical disk spectrum. An empirical relation between the ratio of X-ray to optical flux and orbital period suggests a period {   $\la$4 hr for a black hole accretor}. Such a short period is expected to yield an asymmetric, precessing disk, and we propose that the observed decay lightcurve  is due to modulation of the mass transfer rate due to the disk precession. 

\end{abstract}





\section{Introduction}

{  We have been monitoring the central region of M31 with {\em  Chandra}, averaging $\sim$1 observation per month over 1999--2012,} in order to discover X-ray transients. Since then, we have reduced our observation rate to 5 per year.  {  Particularly bright or otherwise interesting} transients are followed up with two {\em HST} ACS observations, the first is taken a few weeks after outburst, and the second observation is normally taken $\sim$6 months later; this allows us to identify the counterpart via difference imaging \citep[see e.g.][ and  references within]{barnard2012b}. We summarized the results of the first 12 transients (labeled T1--T12) found via this effort in \citet{barnard2012b}, and performed this analysis for two further transients: XMM J004243.6+412519 \citep[known as M31 ULX2, ][]{barnard13c}, and CXOM31  004252.457+411631.17 \citep[referred to as T13, ][]{barnard14b}.

 In this work we report our findings for CXOM31 004205.77+411330.43, hereafter referred to as T14. We note that T14 is source number 71 in our variability survey of 528 X-ray sources within $\sim$20$'$ of {  the galaxy nucleus} where we made $\sim$200 new X-ray binary {  (XB)} identifications \citep{barnard14a}; {   the 0.3--10 keV luminosities for these sources ranged over $\sim$1--6000$\times 10^{35}$ erg s$^{-1}$}. The initial outburst of T14 was in 2011 July; however, the second observations was made almost 3 years later, in 2014 June.

X-ray transients within the central region of M31 are most likely to contain a low mass secondary, as the majority of stars there are old {  \citep[see e.g.][]{williams03}}.   Low mass X-ray binaries {  (LMXBs)} may be transient X-ray sources due to instabilities in their accretion disks; the disk has two stable phases (hot and cold), and an unstable intermediate phase--- matter accumulates in the disk in the cold phase, and is rapidly dumped onto the compact object in the hot phase \citep[see e.g.][]{lasota2001}. However, the X-rays produced by accretion from the hot disk prevent the disk from cooling; the X-ray luminosity decays exponentially if the whole disk is ionised, and linearly if only part of the disk is ionised\citep {king98}.

\citet{vp94} found an empirical relation between the  ratio of  X-ray and optical luminosities of Galactic X-ray binaries  and their orbital periods, suggestive that the optical emission is dominated by reprocessed X-rays in the disk; this relation holds over a 10 magnitude range in optical luminosity, and appears to be independent of inclination. Their chosen X-ray band was 2--10 keV. For an irradiated accretion disk with radius $a$, X-ray luminosity $L_{\rm X}$, optical luminosity $L_{\rm opt}$, and temperature $T$, $T^4$ $\propto$ $L_{\rm X}$/$a^2$, while the surface brightness of the disk, $S$, $\propto$ $T^2$ for typical XBs \citep{vp94}. Since $L_{\rm opt}$ $\propto$ $Sa^2$,  $L_{\rm opt}$ $\propto$ $L_{\rm X}^{1/2} a$; also $a$ $\propto$ $P^{2/3}_{\rm orb}$, where $P_{\rm orb}$ is the orbital period. 

\citet{vp94} defined $\Sigma$ = $\left(L_{\rm X}/L_{\rm EDD}\right)^{1/2}\left(P_{\rm orb}/1 {\rm hr}\right)^{2/3}$, choosing   $L_{\rm EDD}$ = 2.5$\times 10^{38}$ erg s$^{-1}$ as a normalizing constant, and found 
\begin{equation}
M_{\rm V} = 1.57(\pm0.24) - 2.27(\pm 0.32) \log \Sigma.
\end{equation}
However, \citet{vp94} sampled a mixture of neutron star and black hole binaries, in various spectral states. A cleaner sample was obtained by A. Moss et al. (2015, in prep), who used only black hole transients at the peaks of their outburst, and found
\begin{equation}
M_{\rm V} = 0.84(\pm0.30) - 2.36(\pm0.30) \log \Sigma.
\end{equation}
We note that these two relations only differ significantly in normalization, caused by black hole X-ray binaries having larger disks than neutron star binaries with the same period. We have period estimates for 14 M31 transients (T1--T12, ULX2, and T13) observed by {\em {\em Chandra}} and {\em HST} \citep{barnard2012b, barnard13b, barnard14b}. {  These period estimates appear to be slightly shorter over all than for Galactic transients, ranging from $<$1 hr to 140$^{+50}_{-40}$ hr, and mostly $<$30 hr \citep{barnard2012b, barnard13b, barnard14b}.}

\section{Observations and data reduction}

We  observed the region containing T14 in 101 {\em Chandra} observations (60 ACIS, 41 HRC).
 We observed T14 in outburst for 5.2 ks with the HST/ACS on 2011 July 21  using the F435W filter and the WFC mode (or ``aperture'') (H1, jbm105010, PI M.  Garcia); the 4.8  ks  observation       in quiescence occurred   on 2014 June 26 (H2, jc6b02010, PI R. Barnard), using the WFC1-CTE aperture. The WFC1-CTE aperture improves the charge transfer efficiency by moving the target to the corner of a detector, allowing point sources to be more circular, but does not effect sensitivity; hence the level of background observed in H2 is unaffected by this change. We also analyzed {  a 2001 June 29 $\sim$60 ks  {\em XMM-Newton} observation containing  a previous outburst} that had sufficient counts for spectral fitting (0109270101, PI K. Mason).

All optical analysis was performed with {\sc pc-iraf} Revision 2.14.1, except where noted. The {\em Chandra} and {\em XMM-Newton}  observations were analyzed with {\sc ciao} version 4.6.3 and SAS version 13.0 respectively; X-ray spectra from both telescopes were treated with {\sc xspec} version 12.8.2b.

\subsection{{\em HST}  analysis}

 Each {\em HST} observation included four flat-fielded (FLT) images, and one drizzled  (DRZ) image. The flat-fielded images are corrected for instrumental effects, but not background subtracted; the total number of counts in each pixel is given. The native ACS resolution is comparable to the FWHM of the PSF \citep{fruchter09}. The drizzled image combines the flat-fielded images, removes any cosmic rays, and subtracts the sky background; it is normalized to  give the number of counts per second per pixel.   We used the DRZ images from H1 and H2 to create a difference image; however, we  used the H2 FLT images for our aperture photometry because the DRZ images sometimes contain pixels with slightly negative values, and this can cause problems when estimating the number of photons in a region.

\subsubsection{Creating a difference image}

 We reprojected the  H2 DRZ  image  into the coordinates of the H1  DRZ image, to produce an accurate difference image. To do this, we first registered the H1 and H2  images to the LGS Field 5 image with {\sc ccmap}, using unsaturated stars that were close to the target. Then, we used the {\sc iraf} task {\sc wregister} to make the pixel orientation of the H2 image match that of the H1 image. We registered the H2 image to the H1 image before mapping to reduce the noise during image subtraction.   The difference image was produced by subtracting H2 from H1 using the {\sc ftools} task  {\sc farith}. 

\subsubsection{Measuring the optical counterpart}

For H1 and H2 we found the number of photons for each FLT image within various extraction regions, yielding $C_{\rm tot}$ photons in total over $T$ seconds. We subtracted the H2 total from the H1 total to get $C_{\rm net}$ source photons.
 We converted this  to Vega $B$ magnitude via
\begin{equation}\label{4s}
B \simeq -2.5 \log\left[ C_{\rm net}/T \right] + ZP,
\end{equation} 
having obtained the zero point ($ZP$ = 25.77) from the ACS Zero Point calculator\footnote{http://www.stsci.edu/hst/acs/analysis/zeropoints/zpt.py}; we see from \citet[][Equation 12 and Table 18]{sirianni05} that the conversion from F435W counts  to B magnitude is within 3$\sigma$ of our ZP for B-V = $-$0.09 \citep[assuming a typical disk spectrum][]{liu01}. We also estimated the 4$\sigma$ upper threshold for source detection from the total photon count from H2 ($C_{\rm tot, H2}$):
\begin{equation}
B \simeq -2.5 \log\left[ 4\times C_{\rm tot,H2}^{0.5}/T \right] + ZP.
\end{equation}

 We can convert from $B$ magnitude to $M_{\rm V}$ via
\begin{equation}\label{conv}
 M_{\rm V} = B + 0.09   -   N_{\rm H}\times\left(1+1/3\right)/1.8\times10^{21}  -    24.47, 
\end{equation}
where $N_{\rm H}$ is the line of sight absorption; this accounts for the difference in B and V magnitudes of a typical accretion disc,  a relationship between B band extinction and measured line-of-sight absorption towards the object, and the distance to M31 \citep[see][and references within]{barnard2012b}.

\subsection{{\em  XMM-Newton} analysis}
The {\em XMM-Newton} observation suffered substantial {  flares in the particle background due to soft Solar protons}. We selected good time intervals by making a 10--12 keV pn lightcurve from the whole image with 100 s binning, {  using standard filters}. { We extracted source source and background spectra from the pn, MOS1 and MOS2 data: source  spectra from a circular region that was optimized by the software {  (17$''$ radius)}, and background spectra from a nearby circular region on the same CCD with no point source {  (35$''$ radius)}. We grouped the source spectra  to give a minimum of 20 counts per bin,  created the corresponding response matrixes and ancillary response files, then fitted the spectra with {\sc xspec}. The pn, MOS1, and MOS2 spectra were fitted simultaneously with constants of normalization to account for differences in detector responses.}

{  We obtained uncertainties in the best fit parameters by simulating 1000 spectra with the {\sc multifake} command, with deviations in simulated spectra being drawn from the properties of the observed spectrum. The best fit parameters for these spectra were each ranked in ascending order, with 1$\sigma$ uncertainties derived from the 160$^{th}$ and 840$^{th}$ data points.}

%---------------------------- Fig 1 ------------------------------------------

\begin{figure*}
\includegraphics[scale=0.43]{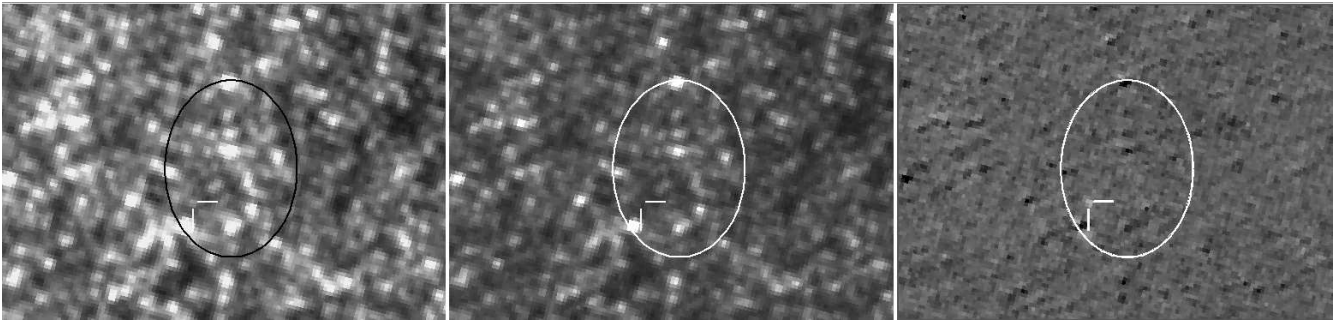}
\caption{{\em Left:} Detail of the 2011 July 21 HST/ACS WFC (H1) image of T14 during outburst. {\em Middle:} 2014 June  (H2) image of T14 in quiescence. {\em Right: } Difference image, where white points were brighter in H1, and dark points were brighter in H2. North is up, East is left.  The 3$\sigma$ uncertainty in X-ray source position (0.9$"$ RA, 1.2$"$ Dec) is indicated by an ellipse. We highlight the counterpart with two lines.  }\label{him}
\end{figure*}
%-----------------------------------------------------------------------------

\subsection{ {\em Chandra}  Analysis}

\subsubsection{Locating the X-ray source}
We used 27 X-ray bright globular clusters (GCs) to register a combined $\sim$350 ks  ACIS image (supplied by Z. Li) to the B band Field 5 image of M31 provided by the Local Galaxy group Survey (LGS) \citep{massey06}. We used {\sc pc-iraf} v2.14.1 to perform the registration, following the same procedure as described in \citet{barnard2012b}. The X-ray and LGS positions of the 27 X-ray bright GCs  were determined using  {\sc imcentroid}; the equivalent FK5 coordinates were calculated for the X-ray and optical position of each GC using {\sc xy2sky} v2.0, distributed with {\sc ftools}. The X-ray positions of each GC were altered to match the LGS positions, allowing the registration of the merged {\em  Chandra} image to Field 5 using the {\sc iraf} task {\sc ccmap}. This registration yielded  1$\sigma$ position uncertainties of  0.11$''$ in R.A., and 0.09$''$ in Dec \citep{barnard2012b}.

 The final uncertainties in the X-ray position of T14 combine the position uncertainties in the X-ray image, and the uncertainties in registering the merged {\em  Chandra} image to the M31 Field 5 LGS image.

{  \subsubsection{Spectral fitting}

Even the best {\em Chandra} ACIS spectrum for T14 yielded few counts. We modeled the spectrum in XSPEC in two ways: i) using a small number of bins with $\ge$20 counts per bin and $\chi^2$ statistics; and ii) using a larger number of bins with $\ge$5 counts per bin and Cash statistics. We estimated the uncertainties for each spectral fit using the same method as for our analysis of the {\em XMM-Newton} spectrum.}

\subsubsection{Estimating luminosities}
For each of our {\em  Chandra} ACIS observations  we extracted source and background 0.3--7.0 keV spectra from circular regions with 10$''$ radii. We then created a response matrix using {\sc mkacisrmf}, and obtained an ancillary response file from {\sc mkarf}.  Spectral analysis was performed using {\sc xspec}, giving unabsorbed 0.3--10 keV luminosities for each observation, allowing us to create a long-term luminosity lightcurve.

{  Most of the {\em Chandra} observations of T14 yielded too few photons for spectral fitting. }However, we are able to estimate the 0.3--10 keV flux (and therefore luminosity) by assuming a particular emission model. If this model is a good approximation, then the luminosities of closely-spaced ACIS and HRC observations should agree; however, if the model is inappropriate, then the ACIS and HRC luminosities should be systematically offset, due to differences in  instrumental response. 

For each faint ACIS spectrum we used XSPEC to estimate the unabsorbed 0.3--10 keV  flux equivalent to 1 count s$^{-1}$, assuming  the best fit emission model for the {\em XMM-Newton} pn spectrum of T14. Multiplying this number by the background-subtracted  intensity yields a flux that is corrected for instrumental effects, and we obtain the luminosity by assuming a distance of 780 kpc \citep{stanek98}. It was important to do this for every observation because the roll angle was not constrained, meaning that the source appeared in different parts of the detectors, at various off-axis angles.

For HRC observations, we included only PI channels 48--293,  thereby reducing the instrumental background. We used the WebPIMMS tool to find the unabsorbed luminosity equivalent to 1 count s$^{-1}$, assuming the same emission model as for the ACIS observations.  We created a 1 keV exposure map for each observation, and compared the exposure within the source region with that of an identical on-axis region, in order to estimate the  exposure correction. We multiplied the background subtracted, corrected source intensity by the correction factor to get the 0.3--10 keV luminosity.

\section{Results}

\subsection{Locating and measuring the  optical counterpart}

The uncertainty in X-ray position for T14 is 0.3$"$ in R.A. and 0.4$"$ in Dec \citep{barnard14a}. Registering H1 to the LGS M31 Field 5 image provided by \citet{massey06} yielded r.m.s. offsets of 0.04$"$ in R.A. and 0.02$"$ in Dec. The uncertainties arising from HST are negligible in comparison with the X-ray uncertainties. In Figure~\ref{him} we present details of the HST/ACS WFC F435W images of T14 from H1 (left), and H2 (middle), along with the difference image (right). For each panel, an ellipse represents the 3$\sigma$ uncertainty in X-ray position with respect to the LGS image (0.9$"$ in R.A., and 1.2$"$ in Dec.). For the difference image, white stars were brighter in H1, while dark stars were brighter in H2. We find evidence for a faint optical counterpart at 00:42:05.818 +41:13:30.00, with respect to the LGS Field 5 B band image. {  This is $\sim$8$'$ offset from the centres of the {\em Chandra} images.}

In \citet{barnard14b} we used a circle with 3 pixel radius (3$\times$FWHM) to estimate the flux for T13; however, this was impossible for T14 because this region was severely contaminated by cosmic rays in the H2 image, and the H2 intensity (7.6$\pm$0.08 count s$^{-1}$) was substantially higher than the H1 intensity (5.5 count$\pm$0.04 s$^{-1}$). Extracting counts from a 3$\times 3$ pixel region  yielded very similar result to extraction from a 2$\times$2 pixel region: background-subtracted  intensities of  0.11$\pm$0.02 and 0.10$\pm$0.02 count s$^{-1}$ respectively. All uncertainties in this work are quoted at the 1$\sigma$ level. 

 For H1, the 3$\times$3 pixel region yielded 6688 photons over 5240 seconds, while the 2$\times$2 region yielded 3406 photons; for H2 the 3$\times$3 region produced 4263 photons over 4790 seconds, while the 2$\times2$ region gave 2658 photons.  These results allow us to estimate the B magnitude of T14 to be 28.21$\pm$0.16 in H1 (using Equation 3) , with a 4$\sigma$ detection limit of 28.6 magnitudes measured for H2 from the 3$\times3$ pixel region (Equation 4). 

  The faintest counterpart that we detected previously had an apparent B magnitude of 24.87$\pm$0.09, but we measured 4$\sigma$ upper limits of B $>$28.7 for two other transients; the detection limit is strongly location-dependent, ranging from B $\sim$26 in the bulge to B $\sim$29 further out \citep{barnard2012b, barnard13b}. %We note that the 4$\sigma$ upper limit for T13 was found to be B $>$26.9, with 180,000 photons accumulated over 4.8 ks from an extraction region with 3 pixel radius, and $\sim$60,000 photons in a 3$\times$3 pixel region. 
 We note that the H1 DRZ file, which is background-subtracted, yields an intensity of 0.16 count s$^{-1}$ for the 3$\times$3 region, and 0.10 count s$^{-1}$ for the 2$\times$2 region, which are  consistent with our intensities obtained by subtracting H2 from H1.%; however, the uncertainties for the DRZ file are considerably large (0.4 and 0.3 count s$^{-1}$ respectively), demonstrating improvement in accuracy gained by using the FLT files. 

\subsection{Characterising the X-ray spectrum}

%---------------------------- Fig 2 ------------------------------------------

\begin{figure}
\includegraphics[scale=0.4]{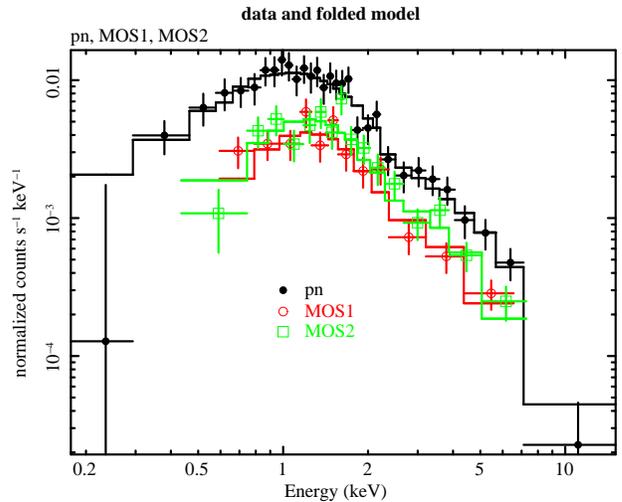}
\caption{Unfolded {\em XMM-Newton} { pn, MOS1 and MOS2 spectra} for T14 from the 2001 June 29 observation of T14, fitted with the best fit absorbed power law model. The y axis shows the model multiplied by channel energy. {  For this model, $N_{\rm H}$ = 2.6$\pm$0.3$\times 10^{21}$ atom cm$^{-2}$ and $\Gamma$ = 1.84$\pm$0.12, $\chi^2$/dof = 42/54; the MOS1 and MOS2 spectra were normalized by factors of 1.02$\pm$0.03 and 1.15$\pm$0.03 respectively in order to account for differences in detector response.   }}\label{spec}
\end{figure}
%-----------------------------------------------------------------------------

\subsubsection{XMM-Newton}

The {\em XMM-Newton} pn spectrum of T14 contained $\sim$550 net source counts, {  with an intensity of 0.0209$\pm$0.0009 count s$^{-1}$}, { while the MOS1 and MOS2 spectra contained 242 and 323 counts respectively}; this is sufficient for simple emission models. {  For the simultaneous spectral fits, $k_1$ and $k_2$ represent the normalizations for MOS1 and MOS2 respectively. }  X-ray binaries exhibit  characteristic emission spectra that depend on the accretor and spectral state. The hard state is common to all X-ray binaries, whether the accretor is a neutron star or black hole \citep{vdk94}; its emission is dominated by inverse Comptonization of cool photons on hot electrons, and may be represented by a power law  with photon index ($\Gamma$) = 1.4--2.1 \citep[see e.g.][]{vdk94,remillard06}. This hard state is observed at luminosities below $\sim$10\% of the Eddington limit \citep{gladstone07,tang11}. Black hole transients often exhibit a thermally dominated state during outburst, characterized by a multi-temperature disk blackbody with inner disk temperature  ($kT_{\rm in}$) $\sim$1 keV \citep{remillard06}. The other states of NS and BH XBs are more complex, with multiple emission components, so we do not attempt to fit them here. 

 { Fitting an absorbed power law model yielded a best fit line of sight absorption ($N_{\rm H}$) of { 2.6$\pm$0.3$\times 10^{21}$ atom cm$^{-2}$} and $\Gamma$ = 1.84$\pm$0.12, $k_1$ = 1.03$\pm$0.03, $k_2$ = 1.15$\pm$0.03; the 0.3--10 keV unabsorbed luminosity was 1.89$\pm$0.12$\times 10^{37}$ erg s$^{-1}$, and $\chi^2$/dof = 42/54. }

{ The best fit disk blackbody model yielded $N_{\rm H}$ = 0.09$\pm$0.03$\times 10^{21}$ atom cm$^{-2}$ and $kT_{\rm in}$ = 1.460$\pm$0.13 keV, $k_1$ = 1.00$\pm$0.03, $k_2$ = 1.15$\pm$0.04;  the 0.3--10 keV unabsorbed  luminosity was 1.25$\pm$0.08 $\times 10^{37}$ erg s$^{-1}$ and $\chi^2$/dof = 56/54.}

The unabsorbed luminosities obtained from these emission models are inconsistent at the 3$\sigma$ level; this is due to the lower column density from the disk blackbody model, caused by the natural downward curvature of the model at low energies. The fit may be improved by using a power law model rather than a disk blackbody model: $\Delta\chi^2$ = 7 for 3 free parameters, and the probability that this improvement is genuine is 93\%, i.e.  not 3$\sigma$. When we consider that the luminosity is consistent with the hard state of any XB (NS or BH accretor), then we find the power law emission model to be most likely. We present the unfolded pn spectrum for T14 assuming the best fit power law emission model in Fig.~\ref{spec}; the y-axis shows the spectrum multiplied by channel energy in order to show the distribution of flux.

\subsubsection{ Chandra}
{ 
The best {\em Chandra} ACIS observation of T14 yielded $\sim$130 counts in the 0.3--7.0 keV range, with no photons detected in the background region over the $\sim$4 ks exposure time. This produced a 6 bin spectrum for $\chi^2$ fitting, and a 25 bin spectrum for fitting with Cash statistics.

The best $\chi^2$ fit absorbed power law model  yielded $N_{\rm H}$ = 8$\pm$6$\times 10^{21}$ atom cm$^{-2}$,  $\Gamma$ = 1.8$\pm$0.5, and a 0.3--10 keV luminosity of 9$\pm$3$\times 10^{37}$ erg s$^{-1}$. The best Cash statistic fit yielded $N_{\rm H}$ = 4$\pm$3$\times 10^{21}$ atom cm$^{-2}$, $\Gamma$ = 1.5$\pm$0.3, and 0.3--10 keV luminosity 7.1$\pm$1.1$\times 10^{37}$ erg s$^{-1}$. These fits are consistent with those obtained from the XMM-Newton spectrum.

Fitting an absorbed  disk blackbody model with  $\chi^2$ statistics resulted in $N_{\rm H}$ = 3$\pm$3$\times 10^{21}$ atom cm$^{-2}$, k$T$ = 2.1$\pm$0.7, and $L$ = 5.2$\pm$0.8$\times 10^{37}$ erg s$^{-1}$. Cash statistics  yielded $N_{\rm H}$ = 2$\pm$2$\times 10^{21}$ atom cm$^{-2}$, and kT = 2.0$\pm$0.7, with a 0.3--10 keV luminosity of 3.1$\pm$0.6$\times 10^{37}$ erg s$^{-1}$.

}
\subsection{The X-ray lightcurve}

%---------------------------- Fig 3 ------------------------------------------

\begin{figure}
\includegraphics[scale=0.4]{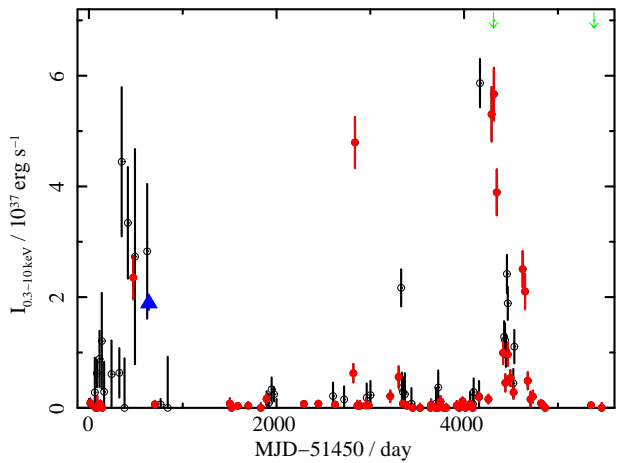}
\caption{Unabsorbed 0.3--10 keV lightcurve for T14, from ACIS (filled circles), HRC (hollow circles) and {\em XMM-Newton} (triangle) observations, assuming an absorbed power law with { $N_{\rm H}$ = 2.6$\times 10^{21}$ atom cm$^{-2}$ and $\Gamma$ = 1.84}.  We also show indicate times of the HST observations with downward arrows.  }\label{lc}
\end{figure}
%-----------------------------------------------------------------------------

We obtained a long-term 0.3--10 keV unabsorbed luminosity lightcurve for T14 from the ACIS and HRC observations by assuming an absorbed power law emission model with { $N_{\rm H}$ = 2.6$\times 10^{21}$ atom cm$^{-2}$ and $\Gamma$ = 1.84}. In Fig.~\ref{lc} we present luminosities from the ACIS (filled circles) and HRC (hollow circles) observations; we indicate the times of the HST observations with downward arrows, and also include the {\em XMM-Newton} observation (triangle).  

We see that T14 exhibited several outbursts; they do not appear to follow the traditional exponential or linear decay seen in Galactic transients \citep{king98}; we discuss a possible reason for this below. The luminosities obtained from {\em XMM-Newton}, ACIS and HRC are all consistent for the first outburst despite the very different instrumental responses, and the ACIS and HRC luminosities appear to be in good agreement for the whole lightcurve, showing that the emission model is probably a good approximation to the true spectrum.

 However, we caution that we did not observe any period where the luminosity exceeded 2$\times$10$^{37}$ erg s$^{-1}$ ($\sim$10\% Eddington for a 1.4 $M_\odot$ NS) with more than one detector. Hence, we cannot determine whether the spectrum evolved as the luminosity increased.

\subsection{Period estimation}

%---------------------------- Fig 4 ------------------------------------------

\begin{figure}
\includegraphics[scale=0.4]{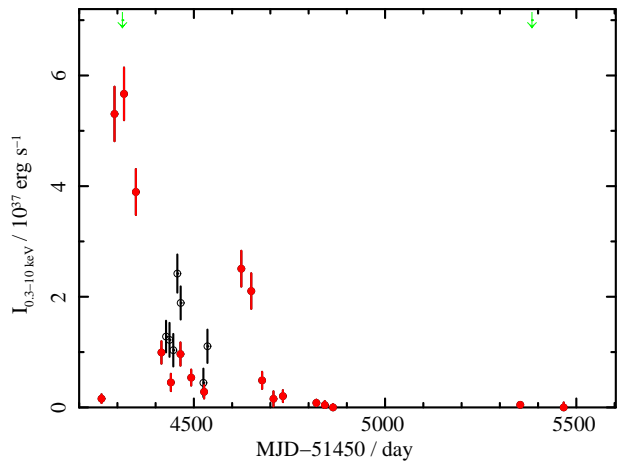}
\caption{Close-up of the outburst examined by {\em Chandra} and HST. The axes and symbols are the same as for Fig.~\ref{lc}.}\label{closeup}
\end{figure}
%-----------------------------------------------------------------------------

Assuming an apparent B magnitude of 28.21 for T14 during H1 and line-of-sight absorption equivalent to 2.6$\times 10^{21}$ H atom cm$^{-2}$, Equation 5 yields an absolute V magnitude of { +1.9} for T14 in outburst. The highest 2--10 keV luminosity observed from T14 during the outburst was 2.7$\pm$0.2$\times 10^{37}$ erg s$^{-1}$ for our assumed emission model.  We estimated the orbital period  using Equations 1 and 2.

 We included the uncertainties in $L_{\rm X}$, $M_{\rm V}$ as well as the uncertainties in the relations themselves by simulating 1000 instances using a Park-Miller random number generator to obtain numbers from 0 to 1, and the Box-Muller technique to convert these numbers into a normal distribution: if we obtain two random numbers $a$ and $b$ between 0 and 1, then $\sqrt{-2\ln a} \cos\left(2\pi b\right)$ and $\sqrt{-2\ln a} \sin\left(2\pi b\right)$ are two independent random numbers  with a normal distribution. For each parameter, we added a normally distributed random number multiplied by the 1$\sigma$ uncertainty  to the best fit value in order to obtain the value for that instance. 

For each empirical relation between X-ray to optical flux ratio and period  (Equation 1 and Equation 2), we obtained 1000 estimates of the period  and ranked these estimates in ascending order; we estimated the mean for each period as ($P_{840}$ + $P_{160}$)/2 and the 1$\sigma$ uncertainty as ($P_{840}$ $-$ $P_{160}$)/2, where $P_{160}$ and $P_{840}$ are the 160$^{th}$ and 840$^{th}$ ranked periods. {   The estimated period from Equation 1 is {  3.0$^{+1.8}_{-1.2}$ hr}, while Equation 2 yields a period of 1.1$^{+0.7}_{-0.4}$ hr; these uncertainties include the uncertainties in the relations described in Equations 1 and 2 as well as uncertainties in X-ray and optical fluxes}.

\section{Summary and conclusions}

We conducted a $\sim$13 year programme of {\em Chandra} observations of the central region of M31, monitoring for outbursts of X-ray transients. {  Transients that were particularly bright or otherwise interesting} were followed up with HST/ACS, using the F435W filter that approximates the B band
 to search for optical counterparts. CXOM31 004205.77+411330.43 (T14) is one such transient; the outburst was observed by HST  in 2011 July, but the follow-up HST observation occurred only recently, in 2014 June. Difference imaging yielded a counterpart with B magnitude 28.21$\pm$0.16.

Modeling an {\em XMM-Newton} spectrum from a previous outburst {  provided a good fit} from an absorbed power law model with { $N_{\rm H}$ =  2.6$\pm$0.3$\times 10^{21}$ atom cm$^{-2}$ and $\Gamma$ = 1.84$\pm$0.12, with $\chi^2$/dof = 42/54.} {   Fitting the best {\em Chandra} ACIS  spectrum yielded results consistent with the {\em XMM-Newton} fits. Furthermore,  using the {\em XMM-Newton} mode} for ACIS and HRC observations yielded consistent results, suggesting that this is a good description of the spectrum, as instrumental differences would result in substantial systematic offsets in luminosity if the spectrum was very different. Using the best fit column density, and assuming a typically flat accretion disk spectrum, we estimated the absolute V band magnitude to be { $\sim$1.9}. 

The highest observed  2--10 keV luminosity during outburst was  { 2.92$\pm$0.06$\times 10^{37}$} erg s$^{-1}$ with our assumed emission model. Using two empirical relations between the ratio of X-ray to optical flux and the orbital period (Equations 1 and 2), we estimate the orbital period of T14 to be {  3.0$^{+1.7}_{-1.2}$ hr and 1.1$^{+0.7}_{-0.4}$ hr} respectively.

We note that Equation 1 was derived from a mixture of NS and BH XBs in several different spectral states, while Equation 2 was obtained from a more selective sample, including only BH XBs in the thermal dominant state. For several of our transients, the resulting periods varied by a factor $>$3; however, for others, including  T14, the periods are consistent within uncertainties \citep{barnard2012b}.

We also note that  relations between optical to X-ray flux ratio and orbital period best apply to the peak of the outburst, because the optical decay rate is systematically a factor $\sim$2 longer than the X-ray decay \citep{chen97}. If the peak was unobserved, then the X-ray to optical ratio would be higher, resulting in a smaller orbital period. 

It is therefore likely that the orbital period for T14 is {   $<$4 hr for a BH accretor}. This short period may explain the rather erratic behaviour during the decay, as shown in Fig.~\ref{closeup}, because accretion may be modulated by the precession of an asymmetric disk. {  Such behaviour was first observed in  cataclysmic variables (CVs) with low mass ratios (and short orbital periods).}

\citet{osaki89} proposed that  the super-outbursts observed in CVs  with low mass ratios (i.e. short periods) are enhanced by a tidal instability that occurs when the outer disk crosses the 3:1 resonance with the secondary; the additional tidal torque causes the disk to elongate and precess, and also greatly enhances the loss of angular momentum (and therefore the accretion rate). The disk precession is prograde in the rest frame, and the secondary repeats its motion with respect to the disk on the beat period between the orbital period and the precession period, a few percent longer than the orbital period. The secondary modulates the disk's viscous dissipation on this period, giving rise to the maxima in the optical lightcurve known as superhumps. Some short period, persistently bright CVs exhibit permanent superhumps \citep{patterson99,retter00}.
We note that \citet{haswell01} found that X-ray binaries with periods $\la$4 hr are likely to have asymmetric, precessing disks. 

The X-ray binary associated with the M31 globular cluster B158 is observed at a high inclination, and exhibits periodic 10017$\pm$50 s ($\sim$2.8 hr) variation in the X-ray lightcurve in some observations \citep{trud02} but not others \citep{barnard06}. The long-term unabsorbed 0.3--10 keV luminosity lightcurve of XB158 varied between $\sim$4 and $\sim$20$\times$10$^{37}$ erg s$^{-1}$; we suggested that this modulation was due to changes in mass transfer during the disk precession cycle \citep{barnard13}. 
In \citet{barnard06} we conducted three dimensional smooth particle hydrodynamical modeling of XB158, and predicted a disk precession period of 29$\pm$1 times the orbital period, or $\sim$81$\pm$3 hr. This motivated us to observe XB158 30 times over  $\sim$30 days with the Swift XRT in order to search for modulations in the intensity over timescales of a few days. We observed approximately sinusoidal variation of the 0.3--10 keV unabsorbed luminosity over $\sim$4--20$\times 10^{37}$ erg s$^{-1}$ on a 5.65$\pm$0.05 day period \citep{barnard15}.

The outburst of T14 studied with {\em Chandra} and HST has a decay profile (Fig.~\ref{closeup}) that is unlike those of Galactic transient XBs, which exhibit exponential decay when the whole disk is irradiated, and linear decay in case of partial irradiation; the decay can change from exponential to linear when the X-ray luminosity is no longer sufficient to irradiate the whole disk \citep{king98}. We propose that the X-ray luminosity may be modulated by variations in mass transfer from an asymmetrical,  precessing disk.

 %\acknowledgments
\section*{Acknowledgments}
 This research
has made use of data obtained from the {\em {\em Chandra}} satellite,
and software provided by the {\em {\em Chandra}} X-Ray Center (CXC). We thank Z. Li for creating the merged {\em Chandra} image used to register the X-ray image.
We also include analysis of public  archival  data from {\em {\em XMM-Newton}}, an ESA
science mission with instruments and contributions directly
funded by ESA member states and the US (NASA) R.B. was 
funded by the  {\em {\em Chandra}} grant GO3-14096X,
along with the {\em HST} grant GO-13111.05-A.  Support for program \#13111 was provided by NASA through a grant from the Space Telescope Science Institute, which is operated by the Association of Universities for Research in Astronomy, Inc., under NASA contract NAS 5-26555. M.R.G. and S.S.M are partially supported by NASA contract  NAS8-03060.
\label{lastpage}
%\email{aastex-help@aas.org}.

%% To help institutions obtain information on the effectiveness of thei telescopes, the AAS Journals has created a group of keywords for telescope

%% in parentheses, after the keyword, but they will not be verified.

%{\it Facilities:} \facility{Swift (XRT)} \facility{{\em XMM-Newton} (pn)}



%\bibliographystyle{aa}
%\bibliography{mnrasm31}

%\end{thebibliography}

%\begin{thebibliography}{}

%\end{thebibliography}

\clearpage

%% the file, uncomment the \plotone, \plottwo, and
%% If you need a layout that cannot be achieved with \plotone or
%% \plottwo, you can invoke the graphicx package directly with the
%% \includegraphics command or use \plotfiddle. For more information,
%% please see the tutorial on "Using Electronic Art with AASTeX" in the
%% documentation section at the AASTeX Web site,
%% http://www.journals.uchicago.edu/AAS/AASTeX.
%%
%% The examples below also include sample markup for submission of
%% supplemental electronic materials. As always, be sure to check
%% the instructions to authors for the journal you are submitting to
%% for specific submissions guidelines as they vary from

%% This example uses \plotone to include an EPS file scaled to
%% 80% of its natural size with \epsscale. Its caption
%% has been written to indicate that additional figure parts will be
%% available in the electronic journal.

\end{document}